\documentclass[structabstract]{aa}
\usepackage{natbib}
\bibpunct{(}{)}{;}{a}{}{,} 

\usepackage[utf8]{inputenc}
\usepackage{geometry}
\usepackage{graphicx}
\usepackage{subfig}
\usepackage{fixltx2e}
\usepackage{float}
\usepackage{caption}

\usepackage{amsmath}

\title{Circum-planetary discs as bottlenecks\\ for gas accretion onto giant planets} \author{Guillaume Rivier
  \inst{\ref{inst1}, \ref{inst2}}  \and Aurélien Crida \inst{\ref{inst1}}
  \and Alessandro Morbidelli \inst{\ref{inst1}}  \and Yann Brouet \inst{\ref{inst1},\ref{inst3}} }

\institute{Laboratoire Lagrange, UMR7293, Universit\'e de Nice Sophia-antipolis/CNRS/Observatoire de la C\^ote d'Azur,
  B.P. 4229, 06304 Nice Cedex 4, France\label{inst1} \and Formation Supa\'ero, Institut Sup\'erieur de l'A\'eronautique et de
  l'Espace, 10 av. Edouard Belin, B.P. 94235, 31400 Toulouse Cedex 4, France\label{inst2} \and Laboratoire
  LERMA/Observatoire de Paris, 61 avenue de l'Observatoire, 75014, Paris, France \label{inst3}}

\date{Accepted 29/10/2012}

\abstract
  {With hundreds of exoplanets detected, it is necessary to
  revisit giant planets accretion models to explain their mass
  distribution. In particular, formation of sub-jovian planets remains
  unclear, given the short timescale for the runaway accretion of
  massive atmospheres. However, gas needs to pass through a
  circum-planetary disc. If the latter has a low viscosity (as
  expected if planets form in ``dead zones''), it might act as a
  bottleneck for gas accretion.}
  {We investigate what the minimum accretion rate is for a planet
    under the limit assumption that the circum-planetary disc is
    totally inviscid, and the transport of angular momentum occurs
    solely because of the gravitational perturbations from the star.}
  {To estimate the accretion rate, we present a steady-state model of
  an inviscid circum-planetary disc, with vertical gas inflow and
  external torque from the star. Hydrodynamical simulations of a
  circum-planetary disc were conducted in 2D, in a planetocentric
  frame, with the star as an external perturber in order to measure
  the torque exerted by the star on the disc. }
  {The disc shows a two-armed spiral wave caused by stellar tides,
    propagating all the way in from the outer edge of the disc towards
    the planet. The stellar torque is small and corresponds to a
    doubling time for a Jupiter mass planet of the order of 5 Myrs.
    Given the limit assumptions, this is clearly a lower bound of the
    real accretion rate.}  
  {This result shows that gas accretion onto a giant planet can be
  regulated by circum-planetary discs. This suggests that the
  diversity of masses of extra-solar planets may be the result of
  different viscosities in these discs.  }

\keywords{Accretion discs - Planets and satellites : formation - Protoplanetary discs - Planet-disc interactions}

\titlerunning{Circum-Planetary discs as bottlenecks for gas accretion onto giant planets}

\authorrunning{G. Rivier \textit{et al.}}

\begin{document}

\maketitle


\section{Introduction}
\label{sec:Introduction}

The detection of exoplanets is probably one of the most striking
discoveries of the past 50 years in astrophysics. Up to now, more than
700 exoplanets have been found. This constantly increasing number
allows a statistical approach to the analysis of exoplanets. The
resulting statistical distributions are then crucial benchmarks for
planetary formation models. For an exoplanet, the two easiest
parameters to determine are the mass and the semi-major axis of its
orbit. Low mass planets are still hard to detect, but it seems that
the mass distribution of planets more than 10 Earth masses is
double-peaked, with a peak close to the mass of Neptune, and another
peak close to the mass of Jupiter \footnote{\tt
  http://exoplanet.eu/}. The first peak is probably an effect of
observational biases as low mass planets are still hard to detect,
but the second peak seems to be real \citep{Mayor_2011}.

The most popular model for giant planet formation is the core
accretion model of \citet{Bodenheimer_1986}. In this model, a solid
core grows, embedded in a gaseous proto-planetary disc. The gas within
the Bondi radius \citep{Bodenheimer_2000} is gravitationally bound to
the core, and forms an envelope.  When the planet reaches a critical
core mass of about $12-16\ M_{\oplus}$ \citep{Pollack_1996,
  Tajima_1997}, the envelope starts to contract quasi-statically and
gas accretion rate increases, for an evolution timescale of a few
million years. This long phase is followed by a runaway gas accretion
stage occurring as soon as both core and envelope masses are
approximately equal, during which quasi-static contraction and thus
gas accretion rate dramatically increase. This leads to an exponential
growth of the planet, in a few $10^5$ yr, which proceeds as long as
there is gas available.

This raises the question\,: what sets the terminal mass of a giant
planet\,? When a planet becomes massive enough (typically a
Saturn mass), it starts to open a gap in the disc around its orbit
\citep{Papaloizou-Lin-1984,Lin_1986,Crida_2006}. However, gap opening
does not necessarily imply that the gasflow onto the planet has
ceased \citep{Artymowicz_1996} because the gap is not totally
empty. Hydro-dynamical simulations
\citep{Bryden_1999,Kley_1999,Lubow_1999, Lubow_2006} show that a
Jupiter-mass planet keeps accreting gas almost as fast as if
gap-opening had not occurred. Accretion stalls by gap-opening only
once a mass in the range of $5-10$ Jupiter masses is achieved. This is
at odds with the masses of the giant planets in the solar system and
of many of the extra-solar giant planets. In particular, the existence
of Saturn-mass planets is a mystery in this scenario.

An often proposed solution \citep[e.g.\,:][]{Thommes_2008} is that the
giant planets form in “dead zones” – regions in the disc with very low
viscosity – and, consequently, open gaps that are much wider and
cleaner than previously thought. These gaps could inhibit accretion
for planet masses of order of a Jupiter mass or less. However, this
idea fails because, as demonstrated by \citet{Crida_2006}, the depth
and width of a gap does not depend just on viscosity but also on the
temperature (i.e. scale height) of the disc. Thus, even in a dead zone
the gap opened by a Jupiter-mass planet should not inhibit
significantly the accretion of gas into the planet's Hill
sphere. Thus, we believe that the question on the terminal mass of
giant planets is still open.

We notice, however, that the \citet{Pollack_1996} model
considered an omnidirectional gas accretion towards the planet\,; yet,
it is physically impossible to reach this state due to angular
momentum conservation of the inflowing gas. In fact, gas must form a
consistent circum-planetary disc (CPD) once the mass of the planet is
higher than about a hundred Earth masses and the planetary atmosphere
has shrunk well inside the planet's Hill radius
\citep{Ward_2010}. This has been confirmed by numerical simulations
\citep[e.g.][]{Ayliffe_2009a,Ayliffe_2009b}. Once a CPD is formed, the
accretion rate of the planet depends on the ability of gas to lose
angular momentum, either by its re-distribution within the disc
(through viscosity) or exchange with external perturbers (e.g. the
star).
  
Most of the previous simulations considered a significant viscosity in
the CPD, comparable to that of the active zones of the circum-stellar
disc. In this case, angular momentum is redistributed very quickly and
the accretion of gas from the CPD onto the planet is extremely fast,
so that the role of the disc can generally be neglected as shown by
\citet{Papaloizou_2005} who assumed a viscosity about $\alpha =
10^{-3}$. However, it is possible that the CPD is in an MRI dead state
as suggested by \citet{Turner_2010}, leading to vanishing turbulence
and a very low viscosity. Indeed, if the circum-planetary disc has a
very low viscosity, then the transport of angular momentum through
this disc can be very inefficient and gas can only accrete onto the
planet at the rate allowed by the removal of angular momentum by
external perturbations, such as the stellar tide. Consequently, the
circum-planetary disc may act as a bottleneck for gas accretion and
dramatically slow down the planet's growth. If, as a result, the
accretion of a planet does not enter the runaway phase but proceeds at
a more regular pace on a timescale comparable to the circum-stellar
disc lifetime (a few million years)
\citep{Haisch_2001,Hillenbrand_2008} then the observed mass spectrum
of the giant planets may be the consequence of the competition between
gas accretion and gas dissipation.

To test whether the idea of a slow accretion rate through a
low-viscosity disc is realistic, we study in this paper the effect of
the stellar tides. We are aware that Reynolds stresses from waves
driven from the circum-solar disc may allow angular momentum losses in
the CPD as well. However, the accretion rate due solely to the solar
tide provides a lower-bound to the real accretion rate for a CPD of
vanishing viscosity. If the result is encouraging ---\,i.e. the
mass-doubling time for a Jupiter-mass planet from solar tides is
longer than the proto-planetary disc lifetime\,--- then the idea of a
low-viscosity CPD as a regulator of the gas accretion rate onto a
giant planet is promising. In this case, future work will have to
evaluate in detail the gas accretion rate with realistic 3D
hydro-dynamical simulations. Thus, our present paper should be
considered as a first step in a long research plan.

The structure of the paper is the following. In Sect.~\ref{sec:model}
we elaborate an idealised semi-analytical model to evaluate the steady
state accretion rate of a planet that is surrounded by an inviscid disc
which undergoes vertical gas inflow and is submitted to a tidal torque
from a distant star. Then we evaluate the two key parameters that
enter in this model. The vertical gas influx rate is estimated in
Sect.~\ref{sec:gas-inflow} from 3D simulations available in the
literature. To compute the stellar torque, in
Sect.~\ref{sec:numerical-simulations} we perform 2D hydrodynamical
simulations of a disc centred on a Jupiter-mass planet. The disc feels
the gravitational perturbation from the star, assumed to be on a
distant, circular orbit. We measure the torque due to the stellar
tides, after a steady-state is reached.  With these two values in
hand, in Sect.~\ref{sec:results} we estimate the planet accretion
rate, expressed as a timescale for the doubling of the mass of a
Jupiter-planet. Finally, we interpret these results and discuss their
possible implications for giant planet formation theories in
Sect.~\ref{sec:conclusion}.

\section{Model}
\label{sec:model}

The tidal torque and the accretion rate are not related in a trivial
way in an inviscid disc. In this section, we present a simple
model to derive an accretion rate on the planet from the tidal torque
that we will measure in Sect.~\ref{sec:results}.

\subsection{Depletion of the CPD}
\label{sec:depletion-cpd}
The torque exerted by the star on a ring of the disc is not entirely
deposited locally in the disc. Only a fraction of the torque is
deposited. This fraction depends on the disc viscosity and tends to
zero for vanishing viscosity \citep{Martin_2011}.  The rest (i.e. the
totality of the torque in the case of an inviscid disc) is passed to
the adjacent rings through pressure effects \citep[][Appendix
  C]{Crida_2006}. If no torque is deposited on a ring, then the gas in
the ring does not lose angular momentum and does not move towards the
central planet.

It would be incorrect, though, to expect that the wave raised by the
star does not promote gas accretion from an inviscid disc. The torque
is transmitted from each annulus of the disc to its internal neighbour
by pressure all the way to the inner edge of the disc (at the boundary
with the atmosphere). This inner annulus, not being supported by
anything interior to it on a Keplerian motion, then loses angular
momentum and falls onto the planet. Once the innermost annulus is
emptied, the same fate occurs to the second innermost annulus and so
forth. Thus, an inviscid disc is depleted from the inside out
as gas is accreted onto the planet’s atmosphere.

More precisely, let us note $X$ the radius of the region of the CPD
that is emptied by the torque over a time $t$. The total orbital
angular momentum of the gas present within $r<X$ is given by\,:
\begin{eqnarray}
  L & = & \int_0^X \Sigma 2\pi r \times r(r\Omega) \,dr\\
  & = & \frac{4\pi}{5} \Sigma \sqrt{G M_p} X^{5/2}
\end{eqnarray}
(assuming the surface density of the gas, $\Sigma$ is independent of
$r$). Here, $ \Omega = \sqrt{\frac{GM_{p}}{r^3}} $ is the angular
speed of the gas in Keplerian motion at a radius $r$, with $M_p$ the
mass of the planet.

By definition of $X$, $L$ should be exactly opposite to the total
angular momentum taken by the star, which is the torque $T$,
accumulated over time\,: $T\times t$. This gives\,:
\begin{equation}
  X(t) = \left(\frac{5}{4\pi}\frac{|T|}{\Sigma \sqrt{G M_p} }\right)^{2/5}t^{2/5}\ .
\label{eq:Xt}
\end{equation}
We see that the radius $X$ increases with time, as expected.

\subsection{Infilling of the CPD}
\label{sec:filling-cpd}

Here, we assume that the CPD is non-viscous. Thus, the CPD doesn't
spread radially, and any gas input at its outer edge just stays
there. However, \citet{Machida_2008} and \citet{Tanigawa_2011} have
shown that vertical inflow onto the CPD is the main source mass of the
CPD. Let us call $C$ this vertical mass flux per surface unit of the
CPD.

Then, the surface density in an initially void region is $\Sigma =
Ct$. Input into Eq.~(\ref{eq:Xt}), this gives\,:
\begin{equation}
X_C = \left(\frac{5}{4\pi}\frac{|T|}{C \sqrt{G M_p} }\right)^{2/5}\ .
\label{eq:XC}
\end{equation}

This implies that there is a radius $X_C$ (independent of time) such
that inside $r<X_C$, depletion due to the stellar torque $T$ and
infilling due to the vertical inflow $C$ exactly balance out. In an
arbitrary time span, all the gas input within a circle of radius $X_C$
around the planet has exactly the orbital angular momentum taken by
the stellar torque. As a consequence, the whole gas of this region
falls onto the planet while the vertical inflow completely refills it
in the same time. Thus, the region within $X_C$ reaches a steady-state
density.

The disc outside $X_C$, however, does not accrete and gas piles
up. This issue will be discussed later in Sect.~\ref{sec:gas-piling}.

\subsection{Accretion rate estimate}
\label{sec:accretion-rate-estim}
We can now write the planet accretion rate as a function of the
torque. The mass flow on the planet is simply the mass flow inside
$r<X_C$\,:
\begin{eqnarray*}
  \dot M_p & = & C \pi {X_C}^2
\end{eqnarray*}
\begin{eqnarray}
 \dot M_p  & = & \pi  \left(\frac{5}{4\pi \sqrt{G M_p}}\right)^{4/5} T^{4/5} C^{1/5}
\label{accr-rate}
\end{eqnarray}
Note that the accretion rate depends weakly on the gas injection flux
$C$, whereas it depends almost linearly on the torque exerted by the
star $T$. This equation is central in this paper, as it links the
stellar torque to the planetary accretion rate.

\subsection{The issue of gas-piling beyond $X_C$}
\label{sec:gas-piling}
  
The model introduced in Sect.~\ref{sec:filling-cpd} raises the issue
of gas-piling in the outer part of the disc. Indeed, if the density in
the outer disc continued to increase, the total torque $T$ exerted
through the wave would increase as well, as it is proportional to
$\Sigma$. This would imply an increase of the accretion radius $X$\,,
and of the accretion rate $\dot M_p$. Moreover, the density in the
outer disc could reach the gravitational instability limit, changing
completely the structure of the CPD and leading to outbursts of
accretion, as suggested by \citet{Martin_2011}.

In reality, though, gas cannot pile-up indefinitely. If the density in
the CPD becomes too large, gas is repelled by pressure and the inflow
from the circum-stellar disc has to stop. Thus, in this section we
estimate the density in the outer CPD, by imposing the pressure
equilibrium between the CPD and the circum-stellar disc (CSD). Our
approach is detailed below.

We assume the CSD is much thicker than the CPD. This, and the fact
that the inflow from the CSD to the CPD is along the vertical
direction \citep{Tanigawa_2011} suggests that the CSD ``weights'' on
top of the surface of the CPD. Then, at equilibrium, the pressure in
the CPD at a height $H_{\rm CPD}$ has to be equal to the pressure of
the CSD at the same height.

Let us now assume a standard Gaussian pressure profile in both
discs\,: $P(z) = P_0 \exp \left(-\frac{1}{2} (\frac{z}{H})^2 \right)$
where $P_0$ is the pressure at the mid-plane (denoted hereafter
$P_0^{\rm CSD}$ and $P_0^{\rm CPD}$ for the CSD and CPD
respectively). At height $H_{\rm CPD}$, the pressure in the CSD is
$P_0^{\rm CSD}$, given our assumption that $H_{\rm CSD} \gg H_{\rm
  CPD}$. In the CPD, $P(H_{\rm CPD})=P_0^{\rm CPD}/\sqrt{e}$.  $P_0$
and $\Sigma$ are linked by the equation of state $P_0 = c_s^2 \Sigma$
(where $c_s=H\Omega$ is the sound speed). The equation $P_{\rm
  CPD}(H_{\rm CPD})=P_{\rm CSD}(0)$ can then be solved for the sole
unknown of the problem\,:
\begin{displaymath}
  \Sigma_{\rm CPD}(r) = \sqrt{e}
  \left( \frac{(H/r_\odot)_{\rm CSD}}{(H/r)_{\rm CPD}} \right)^2 
  \frac{M_\odot}{M_p} \frac{r}{r_\odot} \Sigma_{\rm CSD}  
\end{displaymath}
where $M_\odot$ is the mass of the star and $r_\odot$ the semi-major
axis of the planet.

For $(H/r_\odot)_{\rm CSD}=0.05$ \citep{Pietu_2007}, $(H/r)_{\rm
  CPD}=0.3$ \citep{Ayliffe_2009b}, $M_\odot=10^3M_p$, $r_\odot =
14.4\ R_H$ (see simulation set-up in \ref{sec:code}) and $r =
0.3\ R_H$ which corresponds to the truncation radius of the CPD, as
explained in \ref{sec:numerical-parameters}, one gets $H_{\rm
  CSD}=8\times H_{\rm CPD}$, justifying our assumption. This gives\,:
\begin{equation}
  \Sigma_{\rm CPD}(r) =  \left(\frac{r}{0.31\,R_H}\right) \Sigma_{\rm CSD}  
\label{eq:Sigma_CPD}
\end{equation}
with $R_H$ the Hill radius of the planet\,: $R_H =  r_\odot\left(\frac{M_{p}}{3M_{\odot}}\right)^{1/3}$.


So, we estimate that the surface density of the CPD at $0.3\,R_H$ is
about equal to the local surface density of the CSD, and can not
exceed this value. This value of the density will be used in our
simulations for tidal torque calculation in
Sect.\ref{sec:gas-parameters}.

\section{Gas inflow estimate}
\label{sec:gas-inflow}

To apply the method outlined in Sect.~\ref{sec:model}, we need to
estimate $C$: the flux per unit area of the gas injected onto the
CPD. To do that, we use the results from \citet{Ayliffe_2009b}. Those
simulations give the flux of gas through the Hill sphere ($M_H$, identified
in that paper as the ``planet accretion rate''). For a planet of mass
$M_p = 333\,M_{\oplus}$ (which is consistent with our choice of
parameters) and a locally isothermal disc, \citet{Ayliffe_2009b} find
$\dot M_H = 8 \times 10^{-5}\ \mathrm{M_{\mathrm{Jup}}/yr}$.  Given
that almost all of this flux is vertical \citep{Tanigawa_2011}, we can
set $\dot M_H=C \pi R_{H}^2$, which gives
\begin{equation}
C = 2.54 \times 10^{-5}\ \mathrm{M_{Jup}/yr/R_H^{\ 2}}\ .
\label{eq:C_num}
\end{equation}

\section{Numerical simulations and Torque computation}
\label{sec:numerical-simulations}

The model introduced in Sect.~\ref{sec:model} allows us to derive
analytically an accretion rate onto the planet, given a gas inflow on
the whole disc and an accurate knowledge of tidal effects from the
star, i.e. an estimate of the stellar torque on the
disc. Having evaluated the former in the previous section, here
we describe how we compute the torque numerically. We start by 
reviewing the set-up of our simulations.

\subsection{Simulation set-up}
\label{sec:simulation-set-up}

\subsubsection{Code}
\label{sec:code}

To run simulations, we have used the code FARGO\,\footnote{\tt
  http://fargo.in2p3.fr/} from \citet{Masset_2000a,Masset_2000b},
which is a 2D Eulerian, non self-gravitating, isothermal code, with a
polar grid. FARGO is very suitable for simulations of the interactions
between planets and the proto-planetary disc. Here, we focus on the
environment of the planet. Thus, we have slightly modified FARGO to
adapt it to our case of study.

From now on, the central body will be the planet, while the star will
be the orbiting body around the planet and its disc. Simulations take
place in a planetocentric frame, corotating with the star.  The orbit
of the star is circular and fixed.

The mass unit is the mass of the central body\,: the planet,
$M_p$. The distance unit is $R_H$.  The time unit is
$\frac{1}{\Omega_H}$, where $\Omega_H = \sqrt{GM_{p}/R_H^{\,3}}$ is
the angular speed at $r=R_H$. Note that the star's angular speed is
thus $\Omega_{\odot} \sim \sqrt{\frac{GM_{\odot}}{r_{\odot}^3}} =
\frac{ \Omega_H}{\sqrt{3}}$.

The mass of the star is taken to be $ M_\odot = 1000\ M_p$,
which makes our system scaled like the Jupiter--Sun system. This gives
$r_\odot = 14.4\ R_H$.

The star being out of the grid, we do not apply any smoothing in the
computation of the force exerted by the star onto the disc. This is
actually necessary for an accurate computation of the differential,
tidal forces.

\subsubsection{Gas parameters}
\label{sec:gas-parameters}

In our simulations, we take an initial density profile
$\Sigma(r)=\Sigma_0 \left(\frac{r}{R_H}\right)^{-\alpha}$,
with $\Sigma_0 = 5\times 10^{-4}\ M_{p}/R_H^2$.
This is in agreement with the global simulations from
\citet{Ayliffe_2009a}. Moreover, the surface density evaluated at $0.3
R_H$ is $\sim 9\times 10^{-4}\ M_{p}/R_H^2$ as
$\alpha=1/2$ (see below).  This should be compared to the surface
density at the location of Jupiter: $\Sigma_{\rm CSD} \approx
10^{-3}\ M_{\mathrm{Jup}}/R_H^2$ in the Minimum Mass Solar
Nebula \citep{Weidenschilling_1977, Hayashi_1981}. Thus, our disc
fulfils the condition of maximum pile-up, that we evaluated in
Sect.~\ref{sec:gas-piling}.

Due to numerical reasons, FARGO does not allow the modelling of a
perfectly inviscid disc. Thus, we take a low but non-zero kinematic
viscosity $\nu = 10^{-5}\ R_H^2\Omega_H$, unless specified otherwise
and adopt a surface density profile $\alpha=1/2$ so that the viscous
torque between adjacent annuli is zero. Notice, moreover, that our
method to compute the gas accretion onto the planet does not use the
actual flow of the gas, but only the shape of the spiral wave
generated by the star (whose structure is detailed in
Sect.~\ref{sec:disc-structure}). Thus, our results should not be
significantly affected by our choice of $\nu$, as we will check in
Sect.~\ref{sec:torque-results}.

The Equation Of State in FARGO is locally isothermal, with
$P={c_s}^2\,\Sigma$, where $c_s$ is the sound speed and $c_s=H\Omega$.
The aspect ratio, $H/r$, is constant in all of our simulations and
independent of $r$.

The $H/r$ of the CPD has been measured to be $0.3-0.4$ in
\citet{Ayliffe_2009a}. Unfortunately, the code FARGO is not stable for
values of the scale height that are so large. Thus, we have run
simulations with $H/r$ values ranging from $0.05$ to $0.2$, in order
to extrapolate the results to $H/r=0.4$. Fortunately, as we will see
below, the cumulated torque at the inner edge of the disc turns out to
be independent of the $H/r$ value, which makes the extrapolation
trivial.

\subsubsection{Numerical parameters}
\label{sec:numerical-parameters}

In several previous studies, simulations have shown that the disc
reaches an equilibrium distribution with a sharp truncation at its
outer edge \citep{ Ayliffe_2009a,Ayliffe_2009b, Martin_2011}. This
truncation radius is about $0.3 - 0.4 \ R_H$ and is likely set by
tidal truncation effects as outlined by \citet{Martin_2011}. Indeed,
we observed ourselves such a truncation radius in simulations with a
grid extended to the Hill radius.

Outside the truncation radius, the motion of the gas relative to the
planet is strongly non-Keplerian, as noticed by
\citet{Ayliffe_2009a}. The interaction between this region and the CPD
is not clearly understood. Therefore, as we just want to measure the
effect of stellar tides on the CPD, we consider in this paper a limit
case where the CPD is completely isolated and we truncate the grid at
$R_{\rm out} = 0.3\ \mathrm{R_H}$. The inner border radius is $R_{\rm
  in}=0.01\ \mathrm{R_H}$\,; for a Jupiter mass planet at $5.2$~A.U., it
corresponds to about $540\,000$ kilometres, which is smaller than the
semi-major axis of Europa.

We choose a logarithmic radial spacing for the grid (the width of a
ring $\delta r$ is proportional to its radius $r$), which is suitable
to have an accurate resolution very close to the planet. The
resolution is taken such that $\delta \theta=\delta r/r < H/5r$ (see
Table~\ref{tab:results}). This resolution is appropriate to resolve
the pressure density wave exerted by the star, as $H$ is the typical
pressure scale length.

\subsubsection{Boundary conditions}
\label{sec:boundary-conditions}

We use a non-reflecting boundary condition for both the inner and the
outer borders of the disc. This prevents density waves from reflecting
at the border of the grid. It is important for an accurate evaluation
of the torque onto the disc through the wave. It also does not allow
gas inflow or outflow and keeps the mean density on the first and last
rings constant with time, which is consistent with our model of an
isolated disc.

\subsection{Torque computation}
\label{sec:torque-computation}

The gravitational torque exerted by the star on each ring of the CPD
is computed as follows \citep[see][]{Martin_2011}.  First, it is
computed on each grid-cell. To the direct torque $T_g$, we have to
subtract an indirect torque $T_i$ due to the acceleration of the frame
(i.e. of the planet).

Denoting by $\vec r = (x_c, y_c)$ the coordinates of a cell of mass
$m_{\rm cell}$, by $\vec r_\odot = (x_\odot, y_\odot)$ the coordinates
of the star and by $\vec d = \vec r - \vec r_\odot = (x_d, y_d)$ the
mutual distance vector, the direct and indirect forces exerted by the
star on the cell write:

$$ \vec f_g = -\frac{G M_{\odot} m_{\rm cell}}{d^2} \, \frac{\vec d}{\Vert \vec d \Vert} $$ 
$$ \vec f_i = -\frac{G M_{\odot} m_{\rm cell}}{r_\odot^2} \, \frac{\vec r_\odot}{\Vert \vec r_\odot \Vert} $$ 
Thus, the gravitational torque exerted from the star on one cell is :  
\begin{eqnarray*}
  \Vert \vec T \Vert & = &\Vert \vec T_g - \vec T_i \Vert \\
  & = & \Vert \vec r \wedge (\vec f_g - \vec f_i) \Vert  \\
  & = & - [x_c (f_g.y_d - f_i.y_\odot) - y_c (f_g.x_d - f_i.x_\odot)]
\end{eqnarray*} 

Then, we compute the azimuthal sum over all the cells of an annulus,
to obtain the gravitational torque felt by one ring. Finally, the
total torque on the disc is the sum of the torques felt by the
individual rings. Thus, it is useful to introduce the notion of {\it
  cumulative torque}, which is the sum of the torques felt from the
first ring to the ring in consideration. The total torque is therefore
the cumulative torque at the last ring. For our purposes in this paper
we consider the outermost ring to be the first and the innermost one
the last (unlike \citealt{Martin_2011}, who adopted the opposite
convention).

\section{Results}
\label{sec:results}

In this section, we describe the results of numerical simulations on
the structure of the CPD and present our estimate of the stellar
torque. Then, we derive results for the accretion rate according to
our analytical model.

\subsection{Disc structure}
\label{sec:disc-structure}

\begin{figure}
  \centering

\includegraphics[width=0.5\textwidth]{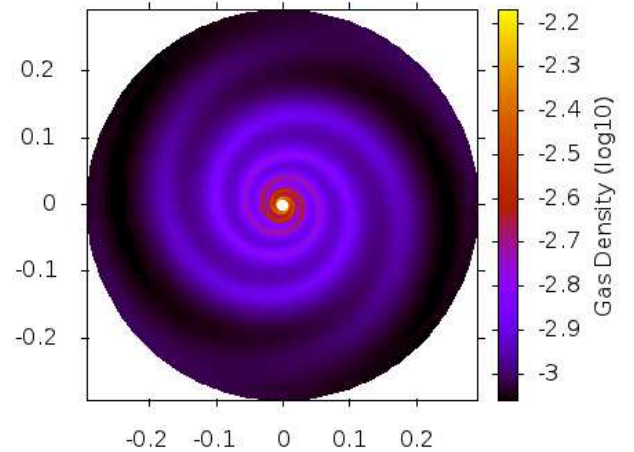}
  \caption{Density map for an aspect ratio of $0.15$\,; the color corresponds to the logarithm of $\Sigma/(M_{\mathrm{Jup}}R_H^{-2})$. The disc is extended from $r = 0.01$ to $0.3\ \mathrm{R_H}$. }
\label{fig:disc-structure}
\end{figure}

As expected, we see a two-armed spiral density wave, created by the
tides of the star. The wave propagates radially inwards with the speed
of sound. As $c_s=H\Omega$, its radial velocity is $H/r$ times its
azimuthal velocity. Thus, the wave has the shape of a logarithmic
spiral, with its pitch angle given by $H/r$. This is clearly seen in
Fig.~\ref{fig:disc-structure}.

\subsection{Torque}
\label{sec:torque-results}

\begin{figure}
  \centering

\includegraphics[width=0.5\textwidth]{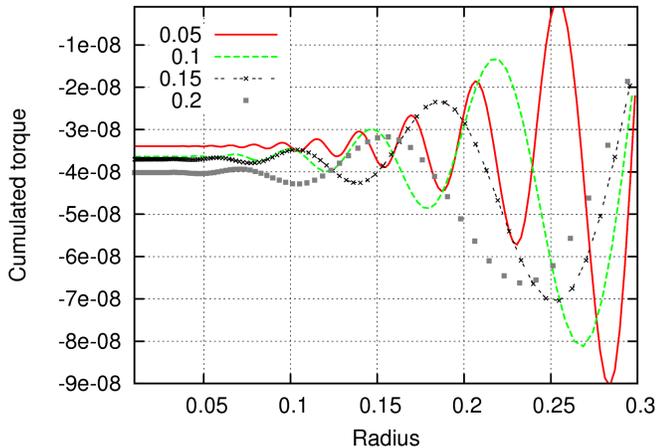}
  \caption{Cumulated torques (in $M_{\rm Jup}R_H^2\Omega_H^2$) as
    a function of radius (in $R_H$) for different aspect ratios.}
  \label{fig:torques}
\end{figure}

We measure the torques after the simulations have reached a
steady-state (after $150$ orbits at $0.3\,R_H$).  \citet{Martin_2011}
showed that the total torque is negative and that most of it is
exerted on the outer region of the disc (which represents less than 10
\% of the disc mass). Therefore, the variations of the cumulative
torque in the inner part of the disc ($r \leq 0.1\ R_H$) are
negligible , as can be seen on Fig.~\ref{fig:torques}. This figure
shows the cumulative torques for various $H/r$, after 250 orbits at
$r=0.3\ R_H$. They oscillate from $r = 0.3$ to $0.1\ R_H$. As
explained by \citet{Martin_2011}, the disc can be divided in four
quadrants, starting from the line connecting the planet to the star,
rotating in the anti-clockwise direction. When the two-arm wave is in
the first and third radiant, the torque is negative\,; when it is in
the second and fourth one, the torque is positive. Thus, the
oscillations of the torque correspond to the passage of the wave from
even to odd quadrants and vice versa, due to its spiral shape. The
wave-length of these pseudo-periodic oscillations is the radial
propagation of the wave during half a revolution around the planet,
that is $\pi\,r\,(H/r)$.

Notice that, although the radial profile of the cumulative torque is
different from simulation to simulation given that the wavelength of
its oscillation is a function of $H/r$, the limit value at the inner
edge of the disc is remarkably insensitive on $H/r$. Thus we can
safely assume that the total cumulative torque for a $H/r=0.3-0.4$
disc (the realistic value of the scale height according to
\citealt{Ayliffe_2009a} is also $\sim 4\times 10^{-8}$.

We can now check whether our choice of the disc viscosity $\nu$
impacts significantly the total torque that we measure. For this
purpose, Table~\ref{tab:results} shows the torques measured in
simulations with various viscosities. Some differences are visible,
which is a sign that our simulations are not dominated by numerical
viscosity.  However, the total torques are very similar to each other
(within $10\%$ for an order of magnitude change in $\nu$). This gives
confidence that the total torque that we estimate is valid also in the
limit of an inviscid disc.

\subsection{Accretion rate}
\label{sec:accr-rate}

In Sect.~\ref{sec:gas-inflow}, we obtained a value for the influx of
gas onto the CPD $C$. Then, thanks to numerical simulations, we
measured the stellar torque for different aspect ratios and low
viscosities and obtained its order of magnitude in
Sect.~\ref{sec:torque-results}. Given these values, we can derive
analytically the planet accretion rate $\dot M$ from
Eq.~\eqref{accr-rate}. The resulting values are listed in
Table~\ref{tab:results}, along with results on the torques and the
corresponding steady-state radius $X_C$. We find that $X_C$ is of the
order of $\sim 0.05\ R_H$, approximately 50 times the current physical
radius of Jupiter.

As can be seen in Table~\ref{tab:results}, the accretion rate hardly
depends on the aspect ratio, and only increases by about $5\%$ when
$H/r$ doubles. This is because, as we said above, the total cumulative
torque is insensitive to the $H/r$ value of the disc. The viscosity of
the gas in the simulation also has very little influence on the
measured torque. In all the cases, we find that the accretion rate is
of the order of\,:
\begin{equation*}
\dot M_p \approx 2 \times 10^{-7}\ M_{\rm Jup}/\mathrm{yr}\ .
\end{equation*}

This accretion rate is low, making Jupiter double its mass in 5 million
years. The torque that we measure is of course proportional to the
surface density of the CPD in the simulations, but we have shown that
the adopted value is realistic, and could hardly be larger. The
parameter $C$ may be poorly constrained, but the dependence of $\dot
M_p$ on $C$ is such that two orders of magnitude difference in $C$
would only change $\dot M_p$ by a factor $3$. Therefore, this
order-of-magnitude estimate of the accretion rate in an inviscid CPD
solely perturbed by the star, is robust. As a consequence, the
accretion of giant planets could be much slower than expected, thus
preventing the runaway accretion phase of the planet.

\begin{table*}
  \centering
  \begin{tabular}[center]{|c|c|c|c|c|c|c|}
    \hline
    $H/r$ & $N_r$ & $N_s$ & $\nu$ & $T$ & $X_C$ & $\dot M_p$ \\
    & & & ($R_H^2\Omega_H$) & ($M_{\rm Jup}R_H^2\Omega_H^2$) & ($R_H$)& ($M_{\rm Jup}$/yr) \\
    \hline
    0.05  &  342  &  632 & $10^{-5}$ & $-3.39\times 10^{-8}$ & $5.11\times 10^{-2}$ & $1.87\times 10^{-7}$ \\
    \hline 
    0.1  &  170  &  314  & $3\times 10^{-5}$ & $-3.76\times 10^{-8}$ & $5.33\times 10^{-2}$ & $2.03\times 10^{-7}$ \\
    0.1  &  170  &  314  & $10^{-5}$ & $-3.64\times 10^{-8}$ & $5.26\times 10^{-2}$ & $1.98\times 10^{-7}$ \\
    0.1  &  170  &  314  & $3\times 10^{-6}$ & $-3.49\times 10^{-8}$ & $5.18\times 10^{-2}$ & $1.92\times 10^{-7}$ \\
    0.1  &  170  &  314  & $10^{-6}$ & $-3.32\times 10^{-8}$ & $5.08\times 10^{-2}$ & $1.84\times 10^{-7}$ \\
    \hline
    0.15  &  114  &  212 & $10^{-5}$ & $-3.70\times 10^{-8}$ & $5.30\times 10^{-2}$ & $2.01\times 10^{-7}$ \\
    \hline
    0.2  &  86  &  160  & $3\times 10^{-5}$ & $-3.88\times 10^{-8}$ & $5.40\times 10^{-2}$ & $2.09\times 10^{-7}$ \\
    0.2  &  86  &  160  & $ 10^{-5}$ & $-4.02\times 10^{-8}$ & $5.47\times 10^{-2}$ & $2.15\times 10^{-7}$ \\
    \hline
  \end{tabular}
\caption{Simulation parameters and results. $N_r$ and $N_s$ are the
  number of radial and azimuthal grid-cells respectively. $\nu$ is the
  viscosity, $T$ the cumulated stellar torque, $X_C$ the equilibrium
  radius between the inner part of the disc in steady-state and the
  outer part. Finally, $\dot M_p$ is the accretion rate onto the planet.}
\label{tab:results}
\end{table*}

\section{Conclusion}
\label{sec:conclusion}

The classic model for giant planet formation \citep{Pollack_1996}
predicts that the final phase of gas accretion occurs in a very fast
runaway mode. Consequently, planets should keep accreting gas until
they are so massive they open very wide gaps in the disc,
which occurs when they reach a mass equal to multiple times the mass
of Jupiter \citep{Bryden_1999,Kley_1999,Lubow_1999, Lubow_2006}.

In this paper we considered that, during the alleged runaway growth
phase, a planet should be surrounded by a circum-planetary disc (CPD)
due to angular momentum conservation of the gas global flow. Thus,
most of the gas that it accretes should have passed through this disc.
The viscosity in the CPD may be very low, if the planet is located
in a dead zone. Therefore, we have investigated whether a non-viscous
CPD could act as a bottleneck for gas accretion. If, consequently, the
gas accretion timescale can become comparable to the timescale of gas
removal from the proto-planetary nebula, the observed large spread in
giant planet masses could stem from the competitions between these two
timescales.

Considering that there is no radial drift in a non-viscous disc, we
have developed a model for a steady-state non-viscous CPD.  The disc
is fed by a vertical gas inflow from the surrounding environment
\citep[as previously observed in 3D numerical simulations,
  see][]{Machida_2008, Tanigawa_2011,Ayliffe_2012}, until a pressure
equilibrium is reached. The surface density of the CPD is thus
analytically determined and found in agreement with previous numerical
simulations \citep{Ayliffe_2009a}.

However, the CPD is perturbed by the star. This results in the
formation of a two-armed logarithmic spiral density wave, that
propagates all the way inwards down to the inner edge of our disc. As
a consequence, a negative torque from the star is deposited in the
very inner regions of the disc, where gas consequently falls onto the
planet. We find that the planetary accretion rate depends almost
linearly on the cumulated stellar torque, and weakly on the vertical
gas inflow (Eq.~(\ref{accr-rate})~).

Running 2D simulations of an isolated disc, in a planetocentric frame,
extended to $0.3$ Hill radius, we have studied the effect of the star
on the disc, and measured the torque. We find that the torque is
negative and basically independent on the disc's aspect ratio. This
allowed us to derive the accretion rate of a giant planet surrounded
by a non-viscous CPD perturbed by the star. We find that the mass
doubling time for a Jupiter-mass planet is about 5~Myr. This timescale
is much longer than that in the runaway accretion mode of
\citet{Pollack_1996}. However, such a low accretion rate is valid in
the limit condition of an inviscid disc. In reality, even in a dead
zone the viscosity is not null, though it is still extremely hard to
estimate quantitatively its value. Furthermore, viscosity, if related
to ionisation, can increase with time as the disc becomes less
optically thick \citep{Turner_2010}. As a consequence, the broad mass
range of observed exoplanets may come from a range of viscosities (or
viscosity evolutions) in their original CPDs.

Moreover, we have only studied the effect of the star on gas accretion
inside the CPD. Several other effects may intervene to make the CPD
lose angular momentum, which need to be investigated in the near
future. In particular, interaction with the gas beyond $0.3\,R_H$ from
the planet and outside the Hill sphere may perturb the flow in the
CPD. The study of such interactions would require further
investigations in global simulations and is beyond the scope of this
paper.

In conclusion, it emerges from this paper that the accretion history
for planets more massive than Saturn \citep[the mass beyond which a
  CPD forms][]{Ayliffe_2009b} may be dominated by the viscous
evolution of the CPD.  We speculate that planets with masses between
Saturn's and a few times Jupiter's may have formed in ``dead zones''
with different levels of low viscosity in their CPDs. Instead, very
massive planets (5-10 Jupiter masses), which reached the mass limit of
the runaway growth process imposed by gap opening in the disc, should
have formed in active zones, where CPDs did not act as bottlenecks to
accretion.

\vspace{0.5cm}

\begin{acknowledgements}
G.R. and Y.B.'s internships at O.C.A. were funded by the PPF OPERA.\\
We thank the CRIMSON team, who manages the mesocentre SIGAMM of the O.C.A., on which most simulations were performed.
\end{acknowledgements}

\vfill

\bibliographystyle{aa}
\bibliography{ref}

\end{document}